\documentclass[a4paper,preprintnumbers,twocolumn,groupedaddress,showpacs,nofootinbib,floatfix,superscriptaddress,aps,10pt]{revtex4-2}
\usepackage{graphicx}
\usepackage{epsfig}
\usepackage{bm}
\usepackage{amsfonts}
\usepackage[utf8]{inputenc}
\usepackage{amssymb}
\usepackage{amsmath}
\usepackage{dcolumn}
\usepackage{latexsym}
\usepackage{amsthm}
\usepackage{framed}
\usepackage{cancel}
\usepackage{subfig}
\usepackage{mathtools}
\usepackage{tablefootnote}
\usepackage{multirow}
\usepackage[breaklinks=true, pdfstartview={FitH}, colorlinks=true, linkcolor=blue, citecolor=red, filecolor=magenta, urlcolor=magenta, anchorcolor=green, linktocpage=true]{hyperref}
\usepackage{nameref}
\usepackage[dvipsnames]{xcolor}

\begin{document}

\title{Circular orbits and accretion disk around a deformed-Schwarzschild black hole in loop quantum gravity}

\author{Kourosh Nozari}
\email[]{knozari@umz.ac.ir}
\affiliation{Department of Theoretical Physics, Faculty of Science, University of Mazandaran,
P. O. Box 47416-95447, Babolsar, Iran}

\author{Sara Saghafi}
\email[]{s.saghafi@umz.ac.ir}
\affiliation{Department of Theoretical Physics, Faculty of Science, University of Mazandaran,
P. O. Box 47416-95447, Babolsar, Iran}

\author{Milad Hajebrahimi}
\email[]{m.hajebrahimi@stu.umz.ac.ir}
\affiliation{Department of Theoretical Physics, Faculty of Science, University of Mazandaran,
P. O. Box 47416-95447, Babolsar, Iran}

\author{Kimet Jusufi}
\email[]{kimet.jusufi@unite.edu.mk}
\affiliation{Physics Department, University of Tetova, Ilinden Street nn, 1200, Tetovo, North Macedonia}

\begin{abstract}

In this paper, we study the motion of neutral and electrically charged particles in the vicinity of a deformed-Schwarzschild black hole inspired by Loop Quantum Gravity (LQG). To examine the motion of an electrically charged test particle, we propose an expression for electromagnetic 4-potential that contains the impacts of loop quantum gravity. This electromagnetic 4-potential satisfies approximately the covariant Maxwell's equations to first order in the loop quantum effects. We explore the effects of the loop quantum correction parameter on the particle geodesics. We investigate the innermost stable circular orbits (ISCOs) for both neutral and electrically charged particles in detail, demonstrating that the loop quantum parameter significantly influences on the ISCO radius, causing it to shrink. Finally, we explore the accretion disk around the loop quantum black hole. We delve into the electromagnetic radiation flux, temperature, differential luminosity, and the spectral luminosity as radiation properties of the accretion disk in detail. We show that the loop quantum correction parameter shifts the profile of the electromagnetic flux and accretion disk temperature towards the central object, leading to a slight increase in these quantities.  

\vspace{12 pt}

\end{abstract}

\maketitle

\section{Introduction}\label{sec01}

Black holes are thought to form when compact, massive objects collapse due to exhaustion of gravity-resisting resources. As a result, they are interesting things with a particularly appealing nature. Recent evidence of gravitational waves \cite{Abbott16a,Abbott16b}, astronomical data from the EHT collaborations \cite{EventHorizonTelescope:2019a,EventHorizonTelescope:2019b}, and the electromagnetic spectrum radiated by an accretion disk surrounding a black hole \cite{frank_king_raine_2002,Yuan:2014gma,2018arXiv181007041N} have proved the existence of black holes. These observations bring up exciting new paths for future research into the outstanding features and gravitational properties of black holes in general relativity and modified gravity theories. In fact black holes, particularly astronomically realistic black holes, can be used as a lab to test various theoretical assumptions by using Multi Messenger Astronomy, including quantum gravity (QG) and other modified gravity theories.

Although observational data suggests the existence of these mysterious massive objects, many fundamental questions remain unresolved, including the singularity within black holes. In gravitation theory and relativistic astrophysics, spacetime singularities with infinite curvature or density, captured researchers' interests. Hawking and Penrose's singularity theorems demonstrate that singularities are unavoidable in generic gravitational collapses. The existence of singularities undermines our understanding of the universe under classical general relativity. Concerns have been raised about naked singularities, which may be detected by outside observers and undermine the prediction capability of classical general relativity. These singularities are not hidden within the black hole event horizon.

A fine gravitational theory is primarily expected to address the issue of spacetime singularities. A theory of quantum gravity must incorporate quantum mechanics with general relativity. Among the different methods approaches to QG proposal, loop quantum gravity (LQG) has shown considerable promise, with major progress achieved (see, for example, \cite{rovelli1995discreteness,ashtekar1997quantum,ashtekar2006quantum,han2007fundamental,ashtekar2011loop,zhang2022first,
zhang2023fermions,AbhishekChowdhuri:2023gvu} and the references therein). Loop quantization of spherically symmetric black holes has provided insights into their quantum nature \cite{chiou2008phenomenologicala,gambini2008black,haggard2015quantum,christodoulou2016planck,ashtekar2018quantum,
zhang2020loop,zhang2022loop,lewandowski2023quantum,husain2022fate,han2022covariant,han2023geometry}. The Schwarzschild black hole's singularity is believed to be resolved through the effects of LQG, though the specific mechanism varies by scheme. More interestingly, the quantum geometry effect can replace the big bang singularity with a non-singular huge bounce \cite{stachowiak2007exact,ashtekar2005quantum}, resulting in more effective black hole models \cite{modesto2006loop,bojowald2018signature,chiou2008phenomenologicalb,battista2024quantum,wang2025dynamical,yang2024gravitational}. The quantum Oppenheimer-Snyder (qOS) model in loop quantum cosmology has led to a novel quantum black hole description \cite{lewandowski2023quantum}. By incorporating a transition area and an inner horizon, this model eliminates the classical Schwarzschild singularity. While the exterior spacetime closely resembles a distorted version of the Schwarzschild solution, the loop quantum-corrected black hole shares many features with its classical counterpart. Studies have investigated its implications for black hole shadows and stability \cite{zhang2023black,yang2023shadow}, though notable deviations arise in asymptotically flat spacetimes. \cite{shao2024scalar} uses a similar strategy to describe a quantum black hole with a positive cosmological constant (qOS-dS). This space-time model has been applied to the LQG-AdS case, and its thermodynamics have been explored \cite{wang2024thermodynamics}. Also, in \cite{shi2024higher} this quantum Oppenheimer--Snyder model for higher-dimensional spacetimes is studied.

In this study we examine the mentioned deformed Schwarzschild black hole inspired by loop quantum Gravity. We study the effects of the loop quantum correction parameter to comprehend the nature of spacetime geometry. In order to do so, we examine the motion of electrically neutral and charged particles around the previously described spacetime geometry that is inspired by LQG. Furthermore, in order to shed light on the characteristics of the loop quantum black hole motivated by LQG and its geometry, we investigate the characteristics of the accretion disk surrounding the black hole.

Accretion disks around black holes are poised to become a primary window into strong-field gravity and extreme spacetime curvature. These luminous structures serve as natural laboratories for probing high-energy processes near black hole horizons. Understanding the underlying spacetime structure is essential, as it directly influences particle motion and observable features such as accretion disk properties and the innermost stable circular orbit (ISCO). Therefore, observational data from thin accretion disks—particularly their thermal emission spectra—can serve as a powerful tool for testing strong-field gravity. The motion of test particles, dynamics, emission spectrum, and structure of accretion disks are all strongly reliant on the spacetime geometry. Alternative gravity theories predict various spacetime metrics, particularly in strong-field regions near compact objects such as black holes or neutron stars. This, in turn, can yield observable variations in accretion disk properties \cite{hou2022image,mukherjee2024constraining,majeed2023dynamics,heydari2023thin,uktamov2024particle,hou2022image,zhang2024imaging,kurmanov2024accretion,Boshkayev:2020kle,guo2020innermost}. Motivated by these considerations, we explore a LQG-inspired black hole solution based on the quantum Oppenheimer-Snyder model. Using the Novikov-Thorne thin-disk framework, we analyze the radiation characteristics and distinctive properties of the accretion disk around this quantum-corrected black hole. Additionally, we investigate how LQG effects modify the ISCO and other key features of black hole spacetime.

The remainder of our paper is organized as follows. In the next section, we present a brief review of the qOS model \cite{lewandowski2023quantum} and the corresponding modified metric inspired by LQG. The motion of neutral test particles is examined in Sec. \ref{sec03}, and in Sec. \ref{sec04}, the dynamics of electrically charged particles surrounding the loop quantum black hole are examined. Section \ref{sec05} focuses on the geometrically thin Novikov--Thorne model for the accretion disk surrounding the loop quantum black hole which is followed by covering the temperature profile, differential luminosity, radiative efficiency, and radiant energy flux over the accretion disk. Our findings have been gathered in Sec. \ref{sec06}. We utilize the $(-, +, +, +)$ signature for the spacetime metric throughout this study.

\section{Deformed-Schwarzschild black hole in LQG}\label{sec02}

The four-dimensional quantum Oppenheimer--Snyder model \cite{lewandowski2023quantum,yang2023shadow} begins with an effective inner spacetime area that contains corrections of loop quantum cosmology (LQC) \cite{ashtekar2006quantum}. The effective interior spacetime area is then matched with an outside spacetime region via junction conditions on their common boundary surface.

In particular, in Ref. \cite{ashtekar2006quantum} authors show that the Ashtekar--Pawlowski--Singh (APS) dust particles move along geodesics while maintaining constant 
$\tilde{r}$, $\theta$, and $\varphi$, forming a dust sphere within the range $0\leq\tilde{r}\leq\tilde{r}_{0}$. Then, authors in Ref. \cite{lewandowski2023quantum} for the qOS model assumes a (pseudo)-static, spherically symmetric metric
\begin{equation}\label{iormetric}
ds^{2}=-(1-F(r))dt^{2}+(1-G(r))^{-1}dr^{2}+r^{2}d\Omega^{2}\,,
\end{equation}
where $d\Omega^{2}=d\theta^{2}+\sin^{2}\theta\,d\varphi^{2}$ is the line element of unite 2-sphere while $F(r)$ and $G(r)$ are two unknown functions of $r$. The coordinates $(t, r)$ describe the exterior region, while $(\tilde{\tau}, \tilde{r})$ are used within the dust region. The junction conditions determine the metric, rather than the field equations. The resulting deformed Schwarzschild solution exhibits a black hole mass gap and an effective energy-momentum tensor, potentially establishing a connection between quantum black hole phenomena and dark matter.

Loop Quantum Cosmology (LQC) approaches have been applied to black hole models; however, their interpretations vary. Some propose a bouncing interior that eliminates the Killing horizon, leading to black hole evaporation. Others suggest that the black hole transitions into a white hole through quantum tunnelling. Certain models preserve the exterior structure while replacing singularities with regular edges, enabling multiple Kruskal-like extensions.

A complementary scenario, known as the quantum Swiss Cheese (qSC) model, considers an empty bubble $(0\leq\tilde{r}\leq\tilde{r}_{0})$ embedded within the APS quantum universe. As the universe contracts, the bubble shrinks, potentially forming a black hole unless its size approaches the Planck scale.

Both models rely on the dust surface at $\tilde{r}=\tilde{r}_{0}$, which is matched across the APS and exterior metrics through the identification $(\tilde{\tau}, \tilde{r}_{0}, \theta, \varphi)\sim(t(\tilde{\tau}), r(\tilde{\tau}), \theta, \varphi)$. Ensuring the continuity of the induced metric and extrinsic curvature determines the functions $F(r)$ and $G(r)$, as well as the trajectory of the dust surface, allowing for the derivation of \eqref{iormetric} as the line element of the deformed-Schwarzschild black hole in LQG (dubbed as loop quantum black hole)
\begin{equation}\label{Met}
\begin{split}
& ds^{2}=-f(r)dt^{2}+f^{-1}(r)dr^{2}+r^{2}d\Omega^{2}\,,\\
& f(r)=1-\frac{2GM}{r}+\frac{\alpha G^{2}M^{2}}{r^{4}}\,,
\end{split}
\end{equation}
where $\alpha=16\sqrt{3}\pi\gamma^{3}\ell^{2}_{P}$ in which $\ell_{P}$ represents the Planck length and $\gamma$ is the Barbero--Immirzi parameter. In Ref. \cite{lewandowski2023quantum} the authors identified a straightforward method to determine the functions $F(r)$ and $G(r)$ (see Appendix A of Ref. \cite{lewandowski2023quantum} for more details). One can see that the metric tensor $g_{\alpha\beta}$ associated with the line element of the loop quantum black hole \eqref{Met} has the following non-zero components
\begin{equation}\label{compsmetten}
\begin{split}
& g_{tt}=-f(r)\,,\quad g_{rr}=f^{-1}(r)\,,\\
& g_{\theta\theta}=r^{2}\,,\qquad\,\,\, g_{\varphi\varphi}=r^{2}\sin^{2}\theta\,.
\end{split}
\end{equation}
It is important to note that the form of the metric in equation \eqref{Met} is determined for
\begin{equation}\label{bounce-alpha}
r>r_{b}=\left(\frac{\alpha GM}{2}\right)^{1/3}\,.
\end{equation}
The dust surface radius $a(\tau)r_{0}$ of Ref. \cite{ashtekar2006quantum} spans the interval $[r_{\rm b}, \infty]$. Consequently, as in Ref. \cite{lewandowski2023quantum}, the functions $F(r)$ and $G(r)$ can be specified arbitrarily for $r<r_{\rm b}$. The parameter $M$ corresponds to the ADM mass of the metric tensor \eqref{Met}.

In Ref. \cite{lewandowski2023quantum}, the authors demonstrated that the global structure of spacetime, as described by Eq. \eqref{Met}, depends on the number of roots of $(1-F(r))$. They introduced the parameter $0<\beta<1$ as
\begin{equation}\label{defbeta}
G^{2}M^{2}=\frac{4\beta^{4}}{\left(1-\beta^{2}\right)^{3}}\alpha\,.
\end{equation}
For $0<\beta<1/2$, we have
\begin{equation}\label{MMin}
M<M_{\rm min}:=\frac{4}{3\sqrt{3}G}\sqrt{\alpha}\,,
\end{equation}
According to \cite{lewandowski2023quantum}, the function $(1-F(r))$ does not have a real root, indicating that the metric \eqref{Met} does not admit any horizon. The global causal structure of the maximally extended spacetime is therefore identical to that of Minkowski spacetime. Thus, the value  
\begin{equation} \label{eq:Min1}
M_{\rm min}=\frac{16\gamma\sqrt{\pi\gamma}}{3\sqrt[4]{3}}\frac{\ell_{P}}{G}\,,
\end{equation}  
serves as a lower bound for black holes predicted by the models discussed in Ref. \cite{lewandowski2023quantum} (see \cite{zhang2022loop,giesel2021nonsingular,husain2022quantum} for compatible results). The minimum mass is approximately equal to the Planck mass. Its precise value depends on the Barbero--Immirzi parameter $\gamma$ of LQG, which is estimated to be around $0.2$ \cite{meissner2004black,domagala2004black}.

Consider $M>M_{\rm min}$, where $1/2<\beta<1$. The function $(1-F(r))$ has precisely two roots
\begin{equation}\label{roothorz}
r_{\pm}=\frac{\beta\left(1\pm\sqrt{2\beta-1}\right)}{\sqrt{(1+\beta)(1-\beta)^{3}}}\sqrt{\alpha}\,,
\end{equation}
This makes $t$ a singular coordinate. Thus, $\beta$ acts as a dimensionless parameter that determines whether quantum effects permit or prevent the formation of an event horizon. It encapsulates the interplay between the black hole mass $M$ and the quantum correction parameter $\alpha$. In Ref. \cite{lewandowski2023quantum} the authors follow the same techniques as for the Reissner--Nordstr\"om (RN) metric to expand the metric tensor \eqref{Met}. They showed that the resulting Penrose diagram has a structure comparable to the RN spacetime.

This black hole's (loop quantum black hole) properties were explored and shown to be stable under scalar and vector perturbations using quasinormal modes. The shadows were studied in Ref. \cite{yang2023shadow}. The spacetime model has been applied to the LQG-AdS example, and its thermodynamics have been explored \cite{wang2024thermodynamics}. Also, in Ref. \cite{shi2024higher} the quantum Oppenheimer--Snyder model for higher-dimensional spacetimes was studied.

We further investigate the particle dynamics for two different cases around the deformed-Schwarzschild black hole in Loop Quantum Gravity. This is what we intend to examine in the next section.

\section{The Motion of neutral particles}\label{sec03}

In the present setup, a test particle's trajectory is determined by the geodesic structure of the loop quantum black holes's spacetime. Using the Lagrangian formalism, we examine the time-like geodesics surrounding the static spherically symmetric loop quantum black hole in this section \cite{shukla2022orbital,hussain2015timelike}.

Since this spacetime is spherically symmetric, the line element \eqref{Met} of the loop quantum black hole associated with the metric coefficient $f(r)$ in Eq. \eqref{Met} is invariant under temporal translation and rotation around the axes of symmetry. Consequently, the loop quantum black hole's spacetime has the following two Killing vectors
\begin{equation}\label{kill}
\begin{split}
& {}^{(t)}\zeta^{\mu}\frac{\partial}{\partial x^{\mu}}=(1,0,0,0)\frac{\partial}{\partial x^{\mu}}=\frac{\partial}{\partial t}\,,\\
& {}^{(\varphi)}\zeta^{\mu}\frac{\partial}{\partial x^{\mu}}=(0,0,0,1)\frac{\partial}{\partial x^{\mu}}=\frac{\partial}{\partial\varphi}\,.\\
\end{split}
\end{equation}
We seek to determine the two conserved (constants) quantities for the test particle's motion in the spacetime that these killing vectors entail. We study the motion trajectory of charged and electrically neutral test particles around the loop quantum black hole.

\begin{figure}[hbt!]
\centering
\includegraphics[width=1\columnwidth]{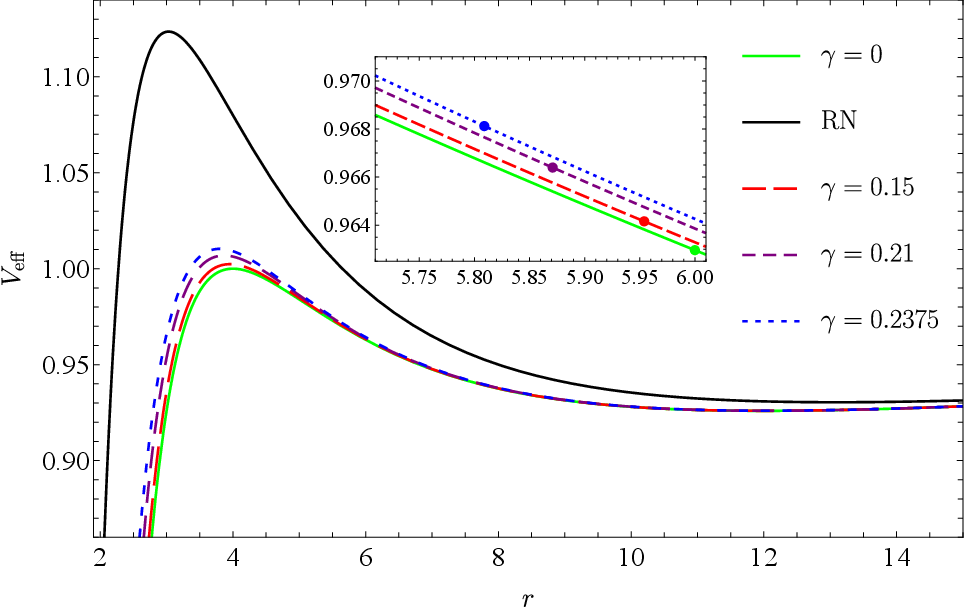}
\caption{\label{Fig1}\small{\emph{$V_{eff}$ of the static spherically symmetric loop quantum black hole shown against $r$ for a range of $\gamma$ values. The Schwarzschild ($\gamma=0$) and RN solutions are shown by the green and black solid lines, respectively. The colored dots show the locations of ISCOs.}}}
\end{figure}

The expression for the Lagrangian of a test particle moving through the loop quantum black hole's spacetime is
\begin{equation}\label{lg}
\mathcal{L}=\frac{1}{2}g_{\mu\nu}\dot{x}^{\mu}\dot{x}^{\nu}\,,
\end{equation}
where a derivative with respect to the affine parameter $\tau$ is represented by an over-dot. The test particle's four-velocity is defined by the formula $\dot{x}^{\mu}\equiv u^{\mu}=(u^{t},u^{r},u^{\theta},u^{\varphi})$. With $\theta=\frac{\pi}{2}$, we are interested in the particle's planar motion on the equatorial plane. Consequently, applying the Euler--Lagrange formula
\begin{equation}\label{EUl}
\frac{d}{d\tau}\left(\frac{\partial\mathcal{L}}{\partial\dot{x}^{\mu}}\right)-\frac{\partial\mathcal{L}}{\partial x^{\mu}}=0\,,
\end{equation}
the two conserved quantities of particle motion that correlate to two Killing vectors can be found as follows
\begin{equation}\label{E}
\frac{dt}{d\tau}=\dot{t}\equiv u^{t}=\frac{E}{f(r)}\,,
\end{equation}
\begin{equation}\label{L}
\frac{d\varphi}{d\tau}=\dot{\varphi}\equiv u^{\varphi}=\frac{L}{r^{2}}\,,
\end{equation}
where the total energy and total angular momentum per unit mass of the particle are denoted by the conserved variables $E$ and $L$, respectively. Additionally, we may determine $\frac{d\theta}{d\tau}=\dot{\theta}\equiv u^{\theta}=0$ using the Euler--Lagrange equation in addition to
\begin{equation}\label{drtau}
\frac{dr}{d\tau}=\dot{r}\equiv u^{r}=\left[-f(r)\left(1-\frac{E^{2}}{f(r)}+\frac{L^{2}}{r^{2}}\right)\right]^{\frac{1}{2}}\,.
\end{equation}
The normalization condition for the test particle's four-velocity, $u^{\mu}u_{\mu}=-1$, and the use of Eqs. \eqref{Met}, \eqref{E}, and \eqref{L} allow one to determine
\begin{equation}\label{rdot}
\dot{r}^{2}=E^{2}-V_{eff}\,,
\end{equation}
where the test particle's effective potential, denoted by $V_{eff}$, can be expressed as follows
\begin{equation}\label{veff}
V_{eff}=f(r)\left(1+\frac{L^{2}}{r^{2}}\right)\,.
\end{equation}
When researching geodesic structure, effective potential analysis is crucial. For instance, the local extremum of the effective potential determines the position of the circular orbits. In Fig. \ref{Fig1}, the behavior of the effective potential $V_{eff}$ for the loop quantum black hole in comparison to the Schwarzschild ($\gamma=0$) and RN examples is shown. As we can see from Fig. \ref{Fig1}, the effective potential increases as the value of the parameter $\gamma$ increases.

A test particle traveling around the black hole at a fixed radial coordinate $r$ is said to be on a circular orbit. Depending on how the effective potential $V_{eff}$ behaves, circular orbits can be either stable or unstable. Accordingly, the primary feature of circular orbits is $\dot{r}=\ddot{r}=0$ (or in other words $u^{r}=\dot{u}^{r}=0$). Therefore, it is possible to confirm that for circular orbits, the conditions $E^{2}=V_{eff}$ and thus, $\frac{dV_{eff}}{dr}=0$ must be met using Eqs. \eqref{drtau} and \eqref{rdot}. Solving these equations simultaneously, the following relations for the total (specific) energy $E$, total (specific) angular momentum $L$, and the angular velocity $\Omega_{\varphi}\equiv\frac{d\varphi}{dt}=\frac{u^{\varphi}}{u^{t}}$ for the test particle in the loop quantum black hole background are obtained by  using Eqs. \eqref{Met}, \eqref{E}, and \eqref{L}
\begin{equation}\label{E2}
E^{2}=\frac{2f^{2}(r)}{2f(r)-rf'(r)}\,,
\end{equation}
\begin{equation}\label{L2}
L^{2}=\frac{r^{3}f'(r)}{2f(r)-rf'(r)}\,,
\end{equation}
\begin{equation}\label{Omega2}
\Omega_{\varphi}^{2}=\frac{1}{2r}f'(r)\,,
\end{equation}
where differentiation with respect to the radial coordinate $r$ is represented by a prime. Eqs. \eqref{E2}-\eqref{Omega2} demonstrate that for unstable circular orbits to exist according to $[2f(r)-rf'(r)>0]$, the total energy and angular momentum must be real.

\begin{figure}[hbt!]
\centering
\includegraphics[width=1\columnwidth]{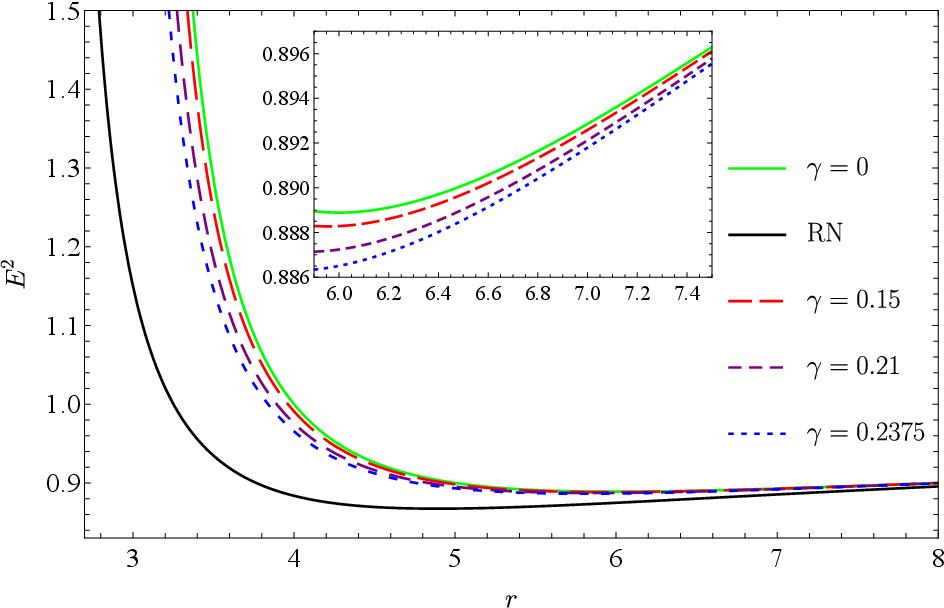}
\caption{\label{Fig2}\small{\emph{The plot of $E^{2}$ of the static spherically symmetric loop quantum black hole versus $r$ for different values of $\gamma$. The green and black solid lines are for the case of Schwarzschild and RN solutions, respectively.}}}
\end{figure}

Figure \ref{Fig2} shows the behaviour of $E^{2}$ versus $r$, indicating that increasing the parameter $\gamma$ decreases the specific energy of the test particle in the spacetime of the loop quantum black hole, while far from the source, the energy becomes almost constant. This figure depicts the energy for the Schwarzschild and RN solutions. For the Schwarzschild case, it consistently has higher values than the loop quantum black hole case while for the RN case, it is smaller than the loop quantum black hole case.

\begin{figure}[hbt!]
\centering
\includegraphics[width=1\columnwidth]{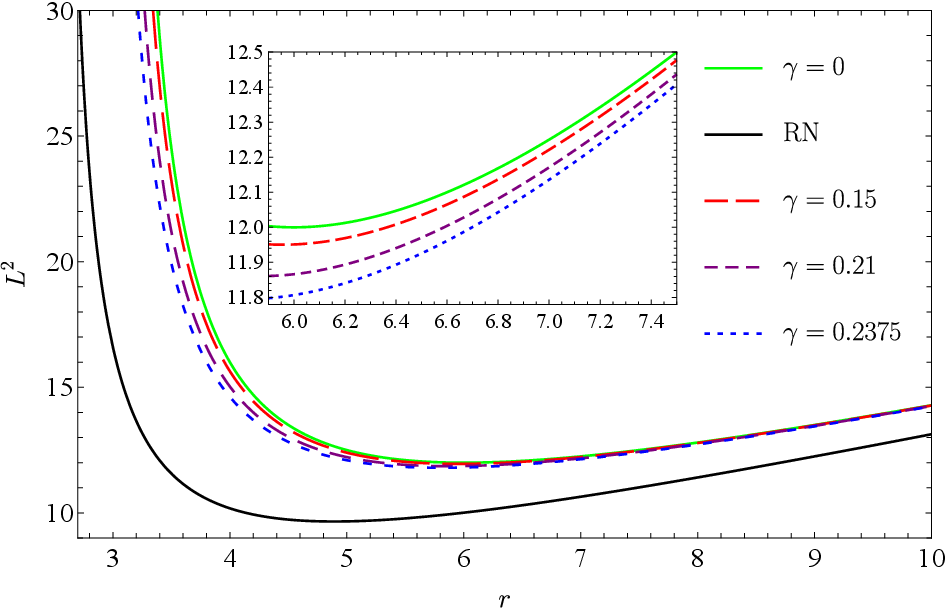}
\caption{\label{Fig3}\small{\emph{The behavior of $L^{2}$ of the static spherically symmetric loop quantum black hole versus $r$ for various $\gamma$ values. The green and black solid lines represent the Schwarzschild and RN solution, respectively.}}}
\end{figure}

Figure \ref{Fig3} shows the relationship between $L^{2}$ and $r$ for the loop quantum black hole in comparison with the Schwarzschild and the RN solutions in GR for various $\gamma$ values. An increase in $\gamma$ results in a decrease in $L^{2}$. Again, for the Schwarzschild case, $L^{2}$ values consistently have higher values than the loop quantum black hole case while for the RN case, they are smaller than the loop quantum black hole case.

\begin{figure}[hbt!]
\centering
\includegraphics[width=1\columnwidth]{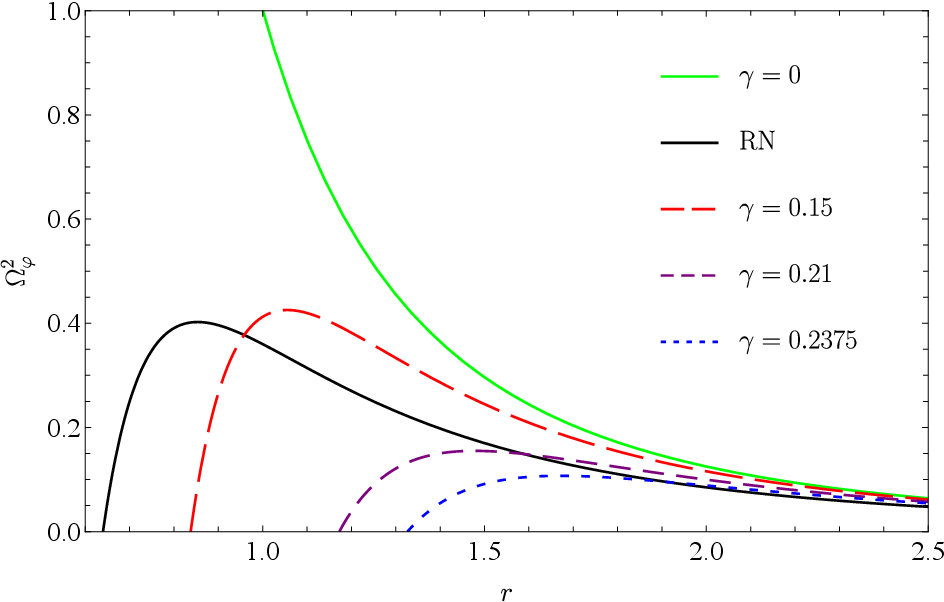}
\caption{\label{Fig4}\small{\emph{The plot of $\Omega_{\varphi}^{2}$ of the static spherically symmetric loop quantum black hole versus $r$ for various $\gamma$ values. The green and black solid lines represent the Schwarzschild and RN solutions, respectively.}}}
\end{figure}

In Fig. \ref{Fig4}, we see the curves of $\Omega_{\varphi}^{2}$ versus $r$ for the loop quantum black hole in comparison with the Schwarzschild and
RN cases. We see that increasing $\gamma$ leads to reduction of the value of $\Omega_{\varphi}^{2}$ so that the curve of the Schwarzschild case contains higher values of $\Omega_{\varphi}^{2}$ than the corresponding ones in the loop quantum black hole.

In addition to fulfilling the conditions of circular orbits, a local minimum of $V_{eff}$ is necessary for a stable circular orbit, hence its second derivative needs to be positive as $\frac{d^{2}V_{eff}}{dr^{2}}>0$ for a stable circular orbit. On the other hand, an unstable circular orbit occurs at a local maximum, meaning $\frac{d^{2}V_{eff}}{dr^{2}}<0$. The boundary (transition) between stable and unstable circular orbits is defined by the innermost (marginally) stable circular orbit (ISCO). The effective potential at this critical radius satisfies
\begin{equation}\label{iscoset}
\frac{dV_{eff}}{dr}=0\,,\qquad \frac{d^{2}V_{eff}}{dr^{2}}=0\,,
\end{equation}
simultaneously. ISCO's existence, $r_{_{ISCO}}$, is entirely a relativistic occurrence. In contrast to classical mechanics, where the effective potential has only one minimum, in general relativity, depending on the choice of $L$, the effective potential can produce either a local maximum and minimum or no extremum. This extremum is associated with a stable outer and an unstable inner circular orbit for the test particle. For a given value of $L$, ISCO is the point at which the stable and unstable circular orbits intersect. The explicit analytical form of ISCO associated with the loop quantum black hole can not be available. Therefore, one can numerically solve the equation set \eqref{iscoset} to yield the values of the ISCO for the test particle moving in the spacetime of the loop quantum black hole. Next, we obtain the numerical values of $r_{_{ISCO}}$, $L_{_{ISCO}}$, and $E_{_{ISCO}}$ for the static spherically symmetric loop quantum black hole for three distinct values of the loop quantum gravity parameter $\gamma$ in Table \ref{Table1}. The ISCO for the Schwarzschild and RN black holes, on the other hand, are $6M$ and $5.4M$. The ISCO associated with the loop quantum black hole therefore decreases as the value of $\gamma$ increases, as Table \ref{Table1} shows.
\begin{table}[hbt!]
  \centering
  \caption{\label{Table1}\small{\emph{The numerical values of $r_{_{ISCO}}$, $L_{_{ISCO}}$, and $E_{_{ISCO}}$ for a test particle moving in the loop quantum black hole spacetime associated with different values of $\gamma$.}}}
  \begin{tabular}{|p{1.5cm}|p{1.5cm}|p{1.5cm}|p{1.5cm}|}
    \hline
    \multicolumn{4}{|c|}{ISCO Parameters} \\
    \hline
    \hline
    ${\gamma}$ & $r_{_{ISCO}}$  & $E_{_{ISCO}}$ & $L_{_{ISCO}}$  \\
    \hline
    \hline
    0 & 6 &  0.94280 & 3.4641 \\
    \hline
    0.15 & 5.95435 & 0.942482 & 3.45694 \\
    \hline
    0.21 & 5.87075 & 0.942206 & 3.44394 \\
    \hline
     0.2375 & 5.80859 & 0.941893 & 3.4344 \\
    \hline
    \end{tabular}
\end{table}

\section{The motion of electrically charged particles}\label{sec04}

Energy and angular momentum can be transferred from the black hole to the accretion disk if a magnetic coupling process is present in the neighborhood of a loop quantum black hole \cite{znajek1976being,blandford1977electromagnetic,wang2002evolution,wang2003magnetic,al2013critical}. Consequently, the magnetic field intensity on the black holes's horizon is given as \cite{hussain2015timelike}
\begin{equation}\label{mgfiho}
B_{h}=\frac{1}{r_{+}}\sqrt{2m_{p}\dot{M}c}\,,
\end{equation}
where $c$ is the speed of light, $m_{p}$ is the magnetization parameter, and the index ($h$) is the abbreviation for horizon. Also, $m_{p}=1$ denotes the equipartition state for the accretion and magnetic coupling process. Black holes and other compact objects can have magnetic fields around them, according to both theoretical and experimental evidence \cite{koide2003magnetic,hussain2014dynamics,jamil2015dynamics,majeed2023dynamics}. Here, we assume that the background geometry is unaffected by the energy of a weak magnetic field \cite{frolov2012weakly}. Weakly magnetized is the term used to describe this kind of loop quantum black hole. Furthermore, we assume that the magnetic field around the black hole is static, axisymmetric, and homogeneous at spatial infinity, where it has the strength $B$. The presence of such a magnetic field is necessary for studying the motion of an electrically charged test particle with mass $m$ and electric charge $q$ around the loop quantum black hole.

In accordance with the methodology presented in Refs. \cite{hussain2015timelike,al2013critical,jamil2015dynamics}, our objective is to compute the magnetic field surrounding the loop quantum black hole. In this setup, according to the above assumptions, we take into account the following Wald-like electromagnetic 4-potential (see \ref{app1} for more details)
\begin{equation}\label{wtemfp0}
A_{\mu}=\left(0, 0, 0, \frac{B}{2}\left(r^{2}+\frac{\alpha}{r^{2}}\right)\sin^{2}\theta\right)\,,
\end{equation}
which approximately satisfies the covariant Maxwell's equations up to first order in $\alpha$. Additionally, the magnetic field as observed by a local observer with 4-velocity $w_{\mu}$ can be defined as follows
\begin{equation}\label{M3}
\mathcal{B}^{\mu}=-\frac{\epsilon^{\mu\nu\lambda\sigma}}{\sqrt{-\bar{g}}}F_{\lambda\sigma}w_{\nu}\,,
\end{equation}
where $\bar{g}=g_{tt}g_{rr}g_{\theta\theta}g_{\varphi\varphi}$ is the metric determinant of the loop quantum black hole described in Eq. \eqref{Met}, $F_{\mu\nu}=\partial_{\mu}A_{\nu}-\partial_{\nu}A_{\mu}$ is the electromagnetic field tensor, and $\epsilon^{\mu\nu\lambda\sigma}$ is the Levi-Cività symbol. The 4-velocity of the observer can be written as follows to satisfy the forward-in-time condition
\begin{equation}\label{M4}
w_{\mu}=\left(\frac{1}{\sqrt{f(r)}}, 0, 0, 0\right)\,.
\end{equation}
Therefore, the magnetic field can be obtained by using Eqs. \eqref{wtemfp0}-\eqref{M4}
\begin{equation}\label{mafifve}
\begin{split}
\mathcal{B}^{\mu} & =\frac{B}{\sqrt{f(r)}}\\
& \times\left(0, \left(1+\frac{\alpha}{r^{4}}\right)\cos\theta, -\left(1-\frac{\alpha}{r^{4}}\right)\frac{\sin\theta}{r}, 0\right)\,,
\end{split}
\end{equation}
in which clearly its $r$- and $\theta$-components are non-zero. At spatial infinity, it is assumed that the magnetic field moves upward along the $\mathrm{z}$-axis \cite{huang2014chaotic}.

\begin{figure}[hbt!]
\centering
\includegraphics[width=1\columnwidth]{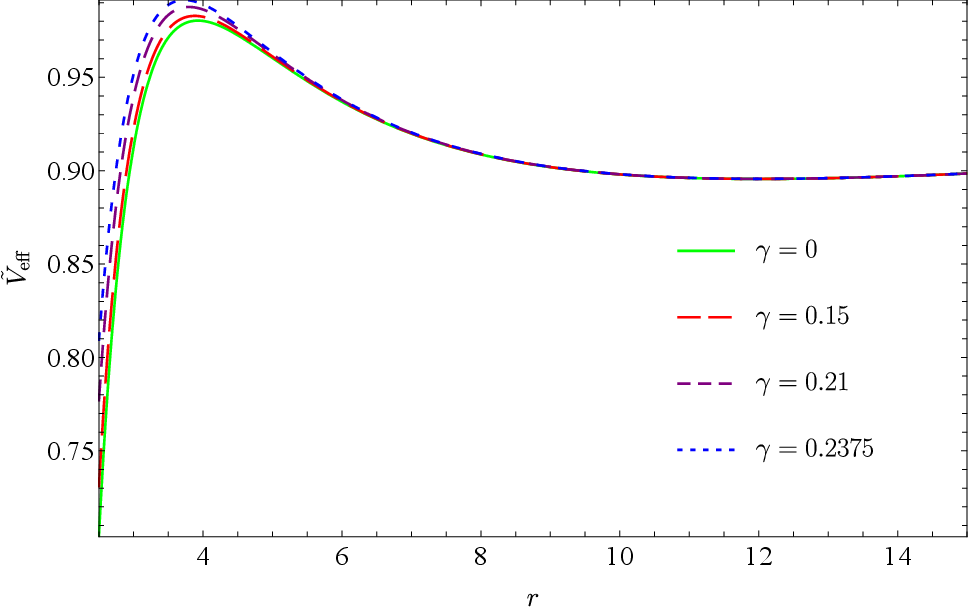}
\caption{\label{Fig7}\small{\emph{The effective potential $\tilde{V}_{eff}$ felt by the massive electrically charged test particle moving in the static spherically symmetric loop quantum black hole's spacetime as a function of $r$ for various $\gamma$ values. The Schwarzschild solution situation in GR is shown by the green solid line.}}}
\end{figure}

The electrically charged test particle moving in the spacetime of the loop quantum black hole has the following Lagrangian
\begin{equation}\label{lagelch}
\tilde{\mathcal{L}}=\frac{1}{2}g_{\mu\nu}\dot{x}^{\mu}\dot{x}^{\nu}+\frac{q}{m}A_{\mu}\dot{x}^{\mu}\,.
\end{equation}
Just like in the last section, we use the Euler-Lagrange equation \eqref{EUl} at the equatorial plane to find
\begin{equation}\label{eelch2}
\dot{t}\equiv\tilde{u}^{t}=\frac{\tilde{E}}{f(r)}\,,
\end{equation}
and
\begin{equation}\label{Lelch2}
\dot{\varphi}\equiv\tilde{u}^{\varphi}=\frac{\tilde{L}}{r^{2}}-\frac{qB}{2m}\left(1+\frac{\alpha}{r^{4}}\right)\,,
\end{equation}
where the electrically charged test particle's four-velocity is $\tilde{u}^{\mu}$, and its specific energy and specific angular momentum are $\tilde{E}$ and $\tilde{L}$, respectively. Once more, the normalizing condition $\tilde{u}^{\mu}\tilde{u}_{\mu}=-1$ can be used on the equatorial plane to obtain
\begin{equation}\label{rdotelch2}
\dot{r}^{2}=\tilde{E}^{2}-\tilde{V}_{eff}\,,
\end{equation}
where $\tilde{V}_{eff}$ is the effective potential felt by the electrically charged test particle, which is defined as
\begin{equation}\label{efpoelch2}
\tilde{V}_{eff}=f(r)\left(1+r^{2}\left(\frac{\tilde{L}}{r^{2}}-\frac{qB}{2m}\left(1+\frac{\alpha}{r^{4}}\right)\right)^{2}\right)\,.
\end{equation}
The plot of $\tilde{V}_{eff}$ versus $r$ for the electrically charged test particle traveling in the spacetime of the loop quantum black hole for various values of $\gamma$ is shown in Fig. \ref{Fig7}. As we can see from this figure, increasing $\gamma$ raises the effective potential values.

As in the previous section, circular orbits require that $\tilde{E}^{2}=\tilde{V}_{eff}$ and $\frac{d\tilde{V}_{eff}}{dr}=0$ be met. Thus, Eqs. \eqref{Met}, \eqref{eelch2}, and \eqref{Lelch2} are used to solve these equations simultaneously and produce the following relations
\begin{equation}\label{Eelch2}
\tilde{E}=\sqrt{f(r)+\frac{f(r)\left[Bqf(r)\left(r^{4}-\alpha\right)+y_{1}\right]^{2}}{m^{2}r^{6}\left(y_{2}\right)^{2}}}\,,
\end{equation}
\begin{equation}\label{Lelch}
\begin{split}
\tilde{L}=\frac{Bqr\left(\alpha+r^{4}\right)f'(r)-4\alpha Bqf(r)-2y_{1}}{2mr^{2}y_{2}}\,,
\end{split}
\end{equation}
\begin{equation}\label{Omegaelch}
\tilde{\Omega}_{\varphi}=\frac{f(r)}{\tilde{E}}\left(\frac{\tilde{L}}{r^{2}}-\frac{qB}{2m}\left(1+\frac{\alpha}{r^{4}}\right)\right)\,,
\end{equation}
where we have defined
\begin{equation}\label{y1}
\begin{split}
y_{1} & =\Big((Bqf(r))^2\left(r^{4}-\alpha\right)^{2}-m^{2}r^{8}(f'(r))^{2}\\
& +2m^{2}r^{7}f(r)f'(r)\Big)^{\frac{1}{2}}\,,
\end{split}
\end{equation}
and
\begin{equation}\label{y1}
y_{2}=rf'(r)-2f(r)\,.
\end{equation}

The massive electrically charged test particle moving in a loop quantum black hole spacetime has an ISCO location that satisfies the requirements \eqref{iscoset}. As we previously stated, the complexity of the metric coefficient \eqref{Met} prevents the availability of the explicit analytical form of ISCO for the electrically charged test particle moving in the loop quantum black hole spacetime. The numerical values of the ISCO for the electrically charged test particle moving in the spacetime of the loop quantum black hole can therefore be obtained by numerically solving the equations set \eqref{iscoset}. This can be done in Table \ref{Table2}, by collecting numerical values of $\tilde{r}_{_{ISCO}}$, $\tilde{L}_{_{ISCO}}$, and $\tilde{E}_{_{ISCO}}$ for various values of $\gamma$ that correspond with the electrically charged test particle for the static spherically symmetric loop quantum black hole. Table \ref{Table2} shows that the ISCO radius of the electrically charged test particle connected with the loop quantum black hole decreases as $\gamma$ increases. By comparing Tables \ref{Table1} and \ref{Table2}, we can see that the ISCO values associated with the loop quantum black hole that is weakly magnetized, are smaller than those associated with the normal loop quantum black hole.
\begin{table}[hbt!]
  \centering
  \caption{\label{Table2}\small{\emph{For an electrically charged test particle moving in the static spherically symmetric loop quantum black hole spacetime, the numerical values of $\tilde{r}_{_{ISCO}}$, $\tilde{L}_{_{ISCO}}$, and $\tilde{E}_{_{ISCO}}$ correspond to various values of $\gamma$ are collected.}}}
  \begin{tabular}{|p{1.5cm}|p{1.5cm}|p{1.5cm}|p{1.5cm}|}
    \hline
    \multicolumn{4}{|c|}{ISCO Parameters} \\
    \hline
    \hline
    ${\gamma}$ & $\tilde{r}_{_{ISCO}}$  & $\tilde{E}_{_{ISCO}}$ & $\tilde{L}_{_{ISCO}}$  \\
    \hline
    \hline
    0 & 5.95709 &  0.96014 & 3.41232 \\
    \hline
    0.15 & 5.91275 & 0.95975 & 3.40531 \\
    \hline
    0.21 & 5.83151 & 0.95905 & 3.39261 \\
    \hline
     0.2375 & 5.77105 & 0.95851 & 3.38328 \\
    \hline
    \end{tabular}
\end{table}

\section{Novikov--Thorne Model of the Accretion Disk: Neutral Particles}\label{sec05}

Black hole accretion disks are essential to astrophysical processes because they are the main sources of electromagnetic radiation emission and mass-energy transfer. A strong relativistic framework for explaining the dynamics and radiative characteristics of geometrically thin, optically thick disks in the vicinity of the loop quantum black hole is offered by the Novikov--Thorne model \cite{novikov1973astrophysics}, one of the several accretion disk models. By adding broad relativistic effects, this model expands on the traditional Shakura--Sunyaev \cite{shakura1973black} prescription and plays a crucial role in comprehending the observational signs of black hole accretion. A thin accretion disk is defined by a vertical scale height $h$ that is much smaller than its radial extension $r$, i.e., $h\ll r$. This characteristic of the thin accretion disk created in the loop quantum black hole's surrounding environment results in negligible vertical pressure and entropy gradients within the disk. As a result, the disk is predominantly influenced by hydrodynamic equilibrium. The accretion disk efficiently radiates heat from its surface, maintaining a low temperature profile. Due to this rapid thermal emission, the energy generated by viscous stresses and dynamical friction cannot accumulate locally. This thermal equilibrium enables the plasma to maintain stable Keplerian motion, forming a geometrically thin accretion disk with a well-defined inner edge at the innermost stable circular orbit (ISCO) around the black hole.

The bolometric luminosity of the accretion disk can be defined as \cite{bokhari2020test, rayimbaev2021dynamics}
\begin{eqnarray}
\mathcal{L}_{bol}=\eta\dot{M}c^2\,,
\end{eqnarray}
Here, $\dot{M}$ denotes the mass accretion rate onto the black hole, while $\eta$ signifies the energy conversion efficiency of the accretion disk. Observational constraints make direct bolometric luminosity measurements challenging, as they depend heavily on black hole characteristics. Theoretical modeling thus becomes essential for estimating this key parameter. A particularly useful approach involves determining the accretion disk's energy efficiency - defined as the maximum extractable energy relative to the rest mass of infalling matter. The energy conversion efficiency is a fundamental parameter in accretion physics, representing the fraction of rest-mass energy that is converted to electromagnetic radiation during the accretion process.  Following \cite{novikov1973astrophysics,page1974disk}, this efficiency can be determined by measuring the photon flux emitted from the disk surface. For a thin accretion disk, the maximum efficiency occurs at the ISCO, where we consider photons that escape to infinity (\cite{narzilloev2022radiation,bardeen1973timelike}) as follows
\begin{equation}
\eta=1-E_{ISCO}\,.
\end{equation}
The radiative efficiency $\eta$ for photons emitted from the accretion disk is determined through our ISCO energy calculations. Using the parameters listed in Table~\ref{Table1}, we analyze the energy efficiency for the loop quantum black hole scenario, with the resulting values presented in Table~\ref{Table3}. Our findings reveal a positive correlation between the quantum parameter $\gamma$ and radiative efficiency. Notably, the $\gamma \rightarrow 0$ limit recovers the Schwarzschild black hole case, where we obtain the characteristic efficiency $\eta \approx 5.72\%$ (consistent with \cite{kurmanov2022accretion}). For comparison, we calculate the Reissner-Nordstr\"{o}m (RN) black hole case to yield $\eta \approx 7.5\%$.
\begin{table}[hbt!]
  \centering
  \caption{\label{Table3}\small{\emph{Value of the radiative efficiency of the accretion disk of the loop quantum black hole arranged for the different values of parameter $\gamma$.}}}
  \begin{tabular}{|p{1.5cm}|p{1.5cm}|p{1.5cm}|}
    \hline
    \multicolumn{3}{|c|}{Radiative efficiency} \\
    \hline
    \hline
    ${\gamma}$ & $r_{_{ISCO}}$   & $\eta$ \\
    \hline
    \hline
    0 & 6 & 5.72\% \\
    \hline
    0.15 & 5.95435 & 5.75\% \\
    \hline
    0.21 & 5.87075 & 5.77\% \\
    \hline
     0.2375 & 5.80859 & 5.81\% \\
    \hline
    \end{tabular}
\end{table}

\subsection{Thin accretion disk properties}\label{sec05ss01}

Here, we investigate the flux generated by the loop quantum black hole's accretion disk. We observe that gas and dust traveling in steady orbits around a black hole or neutron star make up an accretion disk. When the gas and dust begin to move around the central compact object, the background's geometry may cause them to lose energy and angular momentum. Consequently, gas and dust orbits shift to the inner border of the accretion disc and fall toward the direction of the compact object. The gas and dust can heat up and release radiation as a result of this process, which leads to accretion disk radiation. We also anticipate that the loop quantum black hole parameters may have an impact on the accretion disk radiation. Specifically, the gas and dust in the accretion disk surrounding the black hole may be impacted by the loop quantum parameter. It is anticipated that the accretion disk radiation surrounding the loop quantum black hole will be a powerful test to explain important and noteworthy features of the accretion disk and may also have major implications from an observational perspective. The following formula can be used to define the electromagnetic radiation flux \cite{novikov1973astrophysics,shakura1973black,thorne1974disk}
\begin{equation}\label{flux}
\mathcal{F}(r)=\frac{-\dot{M}}{4\pi\sqrt{-g}}\frac{\Omega_{,r}}{(E-\Omega L)^{2}}\int_{r_{_{ISCO}}}^{r}(E-\Omega L)L_{,r}dr\,,
\end{equation}
where the determinant of three-dimensional subspace is given as $g=g_{tt}g_{rr}g_{\varphi\varphi}$.

Here, we suppose that the loop quantum black hole has a total mass of $M=2.5\times 10^{6}M_{\odot}$ and an accretion rate of $\dot{M}=2\times 10^{-6}M_{\odot}\,yr^{-1}$ to analyze the flux distribution of its accretion disk. One can express the rate in terms of the Eddington accretion rate to find $\dot{M}=3.63\times 10^{-4}\dot{M}_{Edd}$, which is within the typical range for supermassive black holes.

The Eddington luminosity represents the maximum energy output an object can achieve when the outward radiation pressure precisely counteracts the inward gravitational pull. Therefore, the Eddington luminosity is given by  
\begin{equation}\label{eddlum}
\mathcal{L}_{Edd}=1.26\times 10^{38}\left(\frac{M}{M_{\odot}}\right)\,erg/s\,,
\end{equation}  
and for an accreting black hole, the Eddington mass accretion rate is defined through the relation \cite{bambi2017black}
\begin{equation}\label{eddmassacc}
\dot{M}_{Edd}=\frac{\mathcal{L}_{Edd}}{\eta c^{2}}\,.
\end{equation}

For a geometrically thin accretion disk with its inner boundary at the ISCO radius, the accretion luminosity typically falls between $5\%$ and $30\%$ of the Eddington limit \cite{mcclintock2006spin,penna2010simulations,heydari2021thin,kulkarni2011measuring}. Based on this, the chosen values of $M$ and $\dot{M}$ remain below the Eddington threshold while characterizing a supermassive black hole. A well-known example is the Sgr A$^{*}$ supermassive black hole, located at the center of the Milky Way, which has an estimated mass of $M=4.1\times 10^{6}M_{\odot}$ and an accretion rate in the range $\dot{M}\sim 10^{-9}-10^{-7}M_{\odot}\,yr^{-1}$ \cite{wang2013dissecting}.

\begin{figure}[hbt!]
\centering
\includegraphics[width=1\columnwidth]{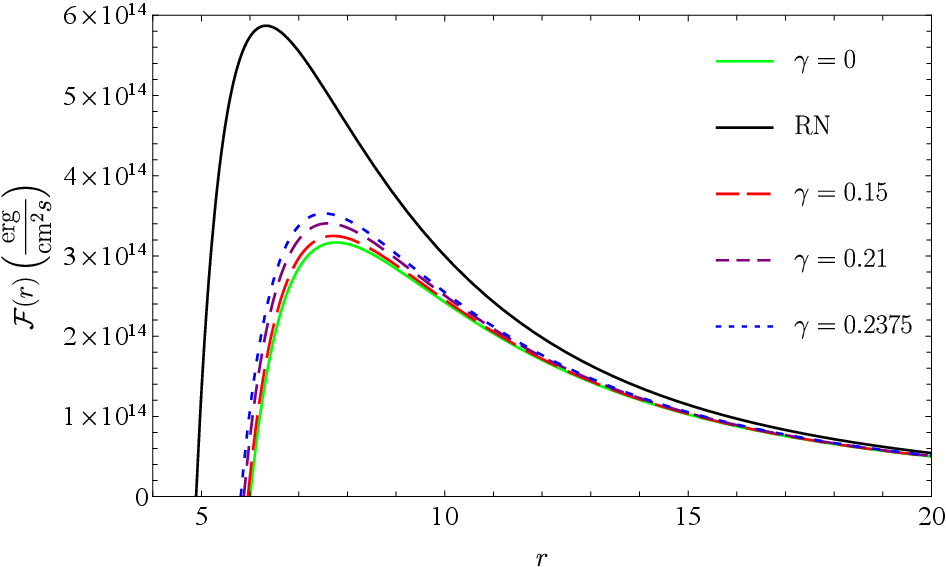}
\caption{\label{Figf}\small{\emph{The illustration of the radial dependence of accretion disk's electromagnetic radiation flux of the loop quantum black hole as the Sgr A$^{*}$ supermassive black hole with $\dot{M}=10^{-7}M_{\odot}\,yr^{-1}$ for various values of $\gamma$. The green thick and black thick curves are associated with the Schwarzschild and RN black hole cases, respectively.}}}
\end{figure}

Eqs. \eqref{E2}-\eqref{Omega2} are also used to further determine the electromagnetic radiation flux. However, it turns out that obtaining the analytical expression of the flow is difficult, thus we use a numerical method, refer to Fig. \ref{Figf}. Figure \ref{Figf} shows that the electromagnetic radiation flux approaches bigger values when the $\gamma$ parameter is increasing. We present the electromagnetic radiation flux for the Schwarzschild and RN black hole cases for a better comparison, in Fig. \ref{Figf}.

\begin{figure}[hbt!]
\centering
\includegraphics[width=1\columnwidth]{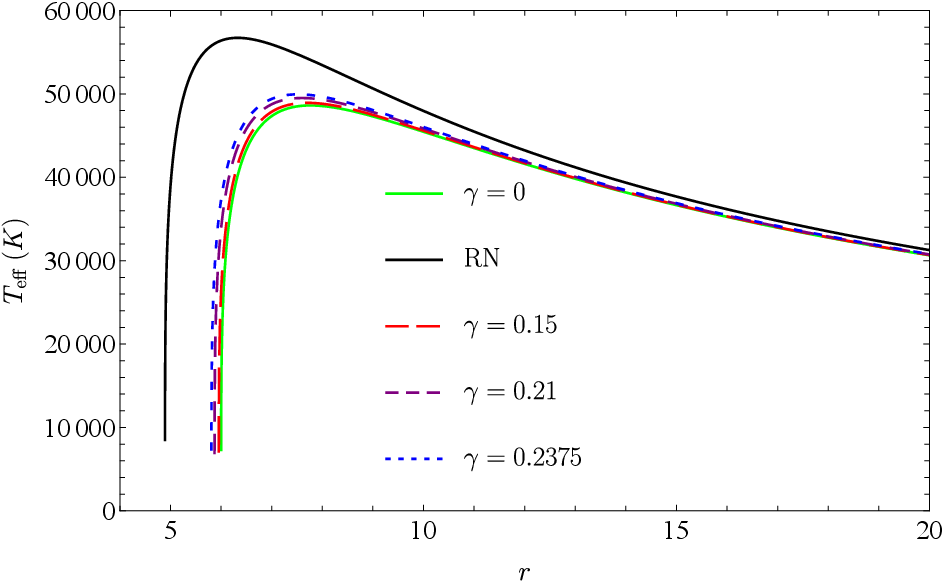}
\caption{\label{Figt}\small{\emph{The illustration of the radial dependence of accretion disk's local effective temperature of the loop quantum black hole as the Sgr A$^{*}$ supermassive black hole with $\dot{M}=10^{-7}M_{\odot}\,yr^{-1}$ for various values of $\gamma$. The green thick and black thick curves are associated with the Schwarzschild and RN black hole cases, respectively.}}}
\end{figure}

As previously described, the Novikov--Thorne model assumes the accreted matter is in thermodynamic equilibrium. This assumption suggests that the disk's radiation can be regarded as ideal black body radiation. Therefore, the energy flux $\mathcal{F}(r)$ and the local effective temperature on the disk $T_{eff}$ are related together according to the Stefan-Boltzmann law as follows
\begin{equation}\label{sblft}
\mathcal{F}(r)=\sigma_{SB}T_{eff}^{4}\,,
\end{equation}
where $\sigma_{SB}=5.67\times 10^{-5}\,erg\cdot cm^{-2}\cdot s^{-1}\cdot K^{-4}$ is the Stefan--Boltzmann constant. Fig. \ref{Figt} illustrates the radial dependency of the local effective temperature on the accretion disk of the loop quantum black hole for different values of $\gamma$. As can be observed from Fig. \ref{Figt}, similarly the local effective temperature on the accretion disk of the loop quantum black hole rises with increasing loop quantum parameter $\gamma$. Also, in Fig. \ref{Figt}, the local effective temperature on the accretion disk of the Schwarzschild and RN black hole cases are also depicted for a better comparison.

\begin{figure}[hbt!]
\centering
\subfloat[\label{temperature2a}\small{\emph{$T_{eff}(r, \gamma)$}}]{\includegraphics[width=0.491\columnwidth]{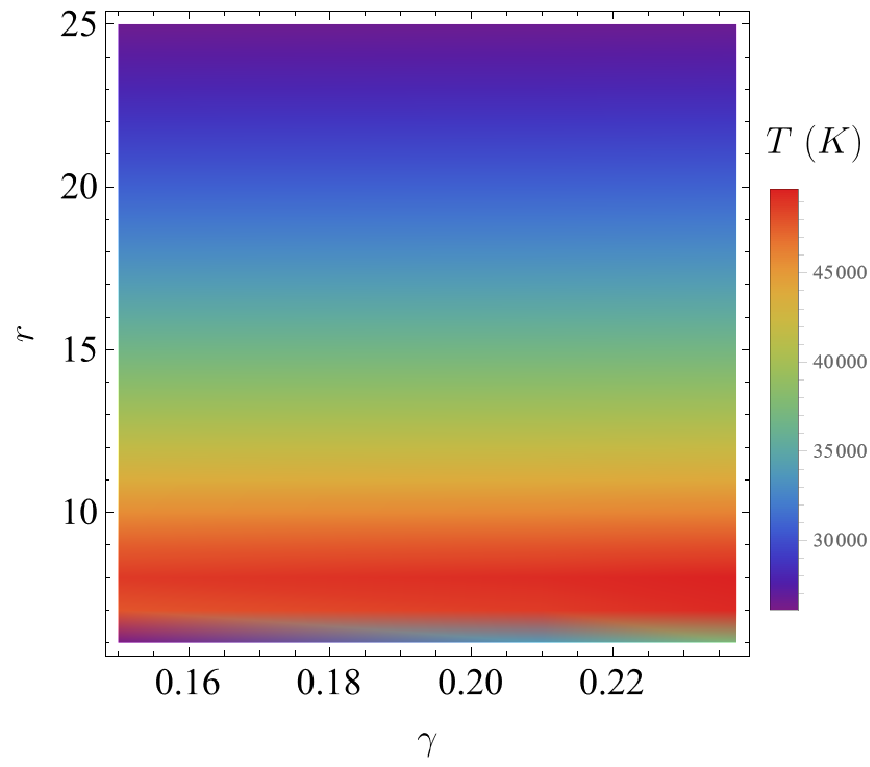}}
\,
\subfloat[\label{temperature2b}\small{\emph{$T_{eff}(r, \gamma=0.2375)$}}]{\includegraphics[width=0.491\columnwidth]{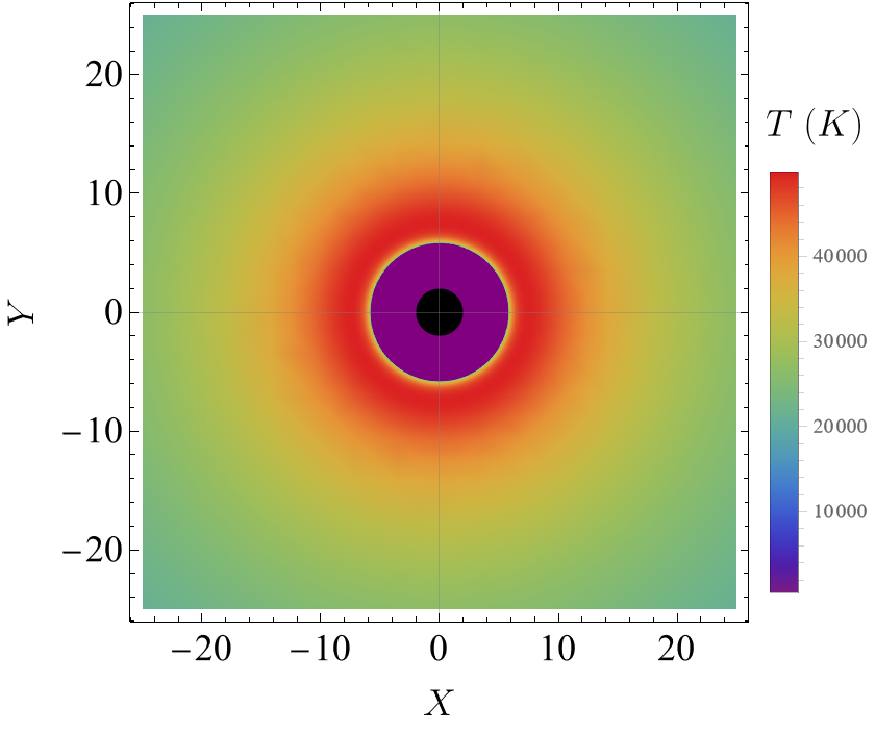}}
\caption{\label{temperature2}\small{\emph{(a): The local effective temperature profile of the loop quantum black hole parameter in terms of $r$ and $\gamma$  in the form of a density plot. (b): The local effective temperature profile on the equatorial plane of $X-Y$ Cartesian coordinates in the form of a density plot.}}}
\end{figure}

In Fig. \ref{temperature2}, we display a density plot of the accretion disk's local effective temperature for more detail. Figure \ref{temperature2} shows how the disk's local effective temperature changes when the loop quantum correction parameter varies. Notably, this density plot demonstrates that the accretion disk's local effective temperature is sufficiently heated, particularly near the black hole, but that it begins to cool toward its outer edge.

\begin{figure}[hbt!]
\centering
\includegraphics[width=\columnwidth]{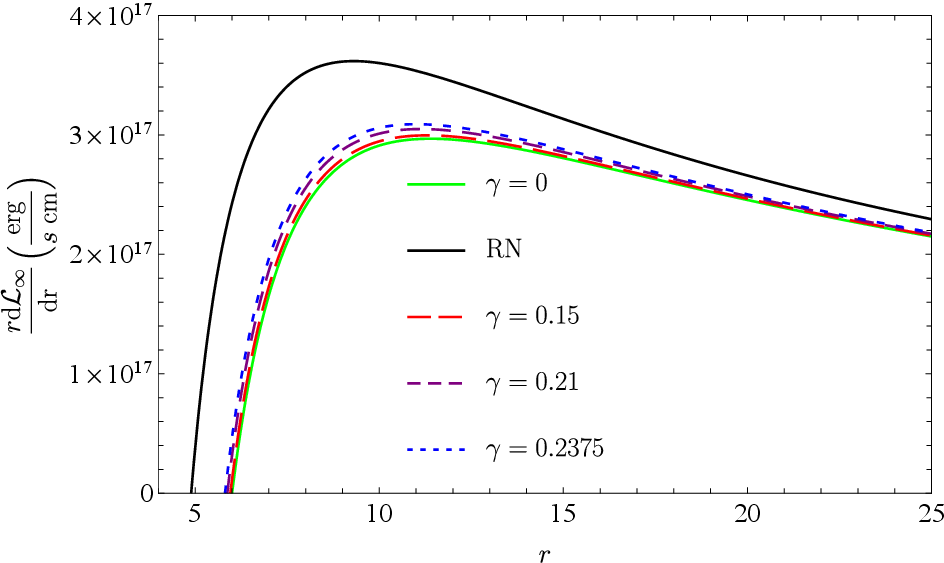}
\caption{\label{Figlu}\small{\emph{The illustration of the radial dependence of accretion disk's differential luminosity received by an observer far from the loop quantum black hole as the Sgr A$^{*}$ supermassive black hole with $\dot{M}=10^{-7}M_{\odot}\,yr^{-1}$ for various values of $\gamma$. The green thick and black thick curves are related to the Schwarzschild and RN black hole cases, respectively.}}}
\end{figure}

The differential luminosity that a distant observer receives at infinity is another important parameter we will discuss. It can be expressed as \cite{novikov1973astrophysics,shakura1973black,thorne1974disk}
\begin{equation}\label{Eq:luminosity}
\frac{d\mathcal{L}_{\infty}}{d\ln{r}}=4\pi r\sqrt{-g}E\mathcal{F}(r)\,.
\end{equation}
We investigate the differential luminosity of the accretion disk and show its radial profile in Fig. \ref{Figlu}. Similar to the accretion disk flux, increasing the loop quantum parameter $\gamma$ elevates differential luminosity. Interestingly, the differential luminosity falls off with increasing distance from the black hole. 
This suggests that the parameter $\gamma$ has a substantial impact in the neighborhood of the black hole. With extensive analysis we indicate that with the influence of the loop quantum correction parameter $\gamma$, the accretion disk can maintain radiative energy and prevent electromagnetic radiation from emitting on its surface.

The change in frequency of a photon as it moves from the source of emission to the observer--that is, its red-shifted black-body spectrum--is associated with the observed luminosity $\mathcal{L}_{\infty}$ of the thin accretion disk around the loop quantum black hole \cite{bhattacharyya2001general,torres2002accretion,banerjee2021implications,karimov2018accretion}. Therefore, the spectral luminosity $\mathcal{L}_{\infty}$ received by a distant observer is as follows
\begin{equation}\label{luminosity2}
\mathcal{L}_{\infty}=\frac{8\pi h\cos\psi}{c^{2}}\int_{r_{i}}^{r_{f}}\int_{0}^{2\pi}\frac{\nu_{e}^{3}rdrd\varphi}{\exp\left[\frac{h\nu_{e}}{k_{B}T_{eff}}\right]-1}\,,
\end{equation}
where $h$ denotes the Planck constant, $k_{B}$ stands for the Boltzmann constant, $\psi$ signifies the inclination angle of the accretion disk, and $r_{i}$ and $r_{f}$ correspond to the inner and outer radii of the disk surrounding the loop quantum black hole. Furthermore, the emitted photon frequency $\nu_{e}$ undergoes redshift to  
\begin{equation}\label{eminu}
\nu=\frac{\nu_{e}}{(1+z)}\,,
\end{equation}
where $\nu$ represents the photon frequency as perceived by a distant observer. Accounting for both gravitational redshift and the Doppler effect, the redshift factor is given by  
\begin{equation}\label{redshiftz}
(1+z)=\frac{1+\Omega r\sin\varphi\sin\psi}{\sqrt{-g_{tt}-g_{\varphi\varphi}\Omega^{2}}}\,,
\end{equation}
and assuming negligible light bending, which may hold as a reasonable approximation for small inclination angles $\psi$, the temperature measured by a remote observer is expressed as \cite{karimov2018accretion,bhattacharyya2001general}  
\begin{equation}\label{tempfarobs}
T_{\infty}=\frac{T_{eff}}{(1+z)}\,.
\end{equation}
Additionally, $\mathcal{L}_{\infty}$ can also represent as the thermal energy flux radiated by the accretion disk encircling the loop quantum black hole.

\begin{figure}[hbt!]
\centering
\includegraphics[width=\columnwidth]{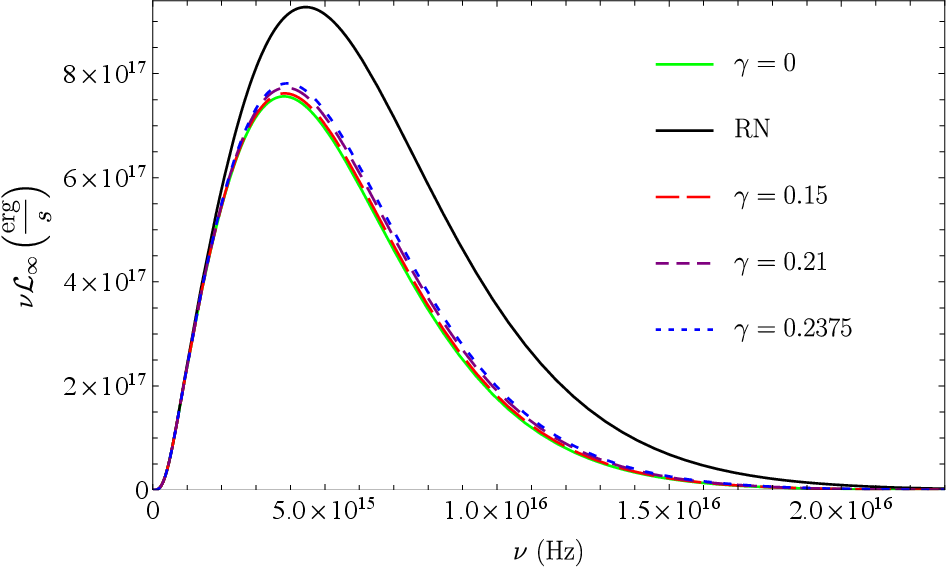}
\caption{\label{Fignulu}\small{\emph{The illustration of the radial dependence of accretion disk's spectral luminosity received by an observer far from the loop quantum black hole as the Sgr A$^{*}$ supermassive black hole with $\dot{M}=10^{-7}M_{\odot}\,yr^{-1}$ for various values of $\gamma$. The green thick and black thick curves are related to the Schwarzschild and RN black hole cases, respectively.}}}
\end{figure}

To plot the spectral luminosity $\mathcal{L}_{\infty}$ of the accretion disk around the loop quantum black hole, we consider $r_{i}=r_{_{ISCO}}$ and $r_{f}=\infty$, as the flux across the disk surface diminishes when $r\to\infty$ in the case of the loop quantum black hole, and also we set $\psi=0$. As a point of note, we used $r_{f}=100$ for simplicity since it is less time-consuming than $r_{f}=\infty$ but does not make much difference in the result. Figure \ref{Fignulu} shows the spectral luminosity received by an observer far from the loop quantum black hole. Similar to the differential flux and temperature, from Fig. \ref{Fignulu}, we see that increasing the loop quantum parameter $\gamma$ leads to amplify the spectral luminosity. The spectrum appears to be a blackbody distribution, with a peak at higher frequencies. When the frequency is low, however, we cannot distinguish between different cases corresponding to each $\gamma$.

\subsection{Observational relevance of LQG-Induced temperature and flux shifts}

Before presenting our final conclusions, it is important to note that the quantum effects in this loop quantum black hole spacetime are governed by the intrinsic Barbero--Immirzi parameter and the minimum mass, raising an intriguing question regarding the potential observational signatures of these effects. Identifying such signatures in current or upcoming experiments could provide a means to directly test or constrain quantum gravitational effects. Over the past decades, extensive studies have explored this possibility using various experimental and observational approaches \cite{yang2024gravitational,Liu:2020ola,brahma2021testing,afrin2023tests,vagnozzi2023horizon,sahu2015gravitational,chen2011complex,cruz2019gravitational,zhu2020observational,
zi2025eccentric}. However, these studies have primarily focused on the classical region outside the black hole horizon, where quantum gravitational effects remain exceedingly small. In this work, we consider a quantum extension of black hole spacetime in which the classical singularity is resolved, leading to quantum effects extending into the exterior region \cite{zhang2023black}. Given this framework, it is of particular interest to investigate how these quantum effects influence the properties of accretion disks, potentially offering new observational avenues to probe quantum gravity. On the other hand, it is important to note that the effects of loop quantum gravity on astrophysical black holes are generally expected to be extremely small. Consequently, the results obtained in this study are most relevant for small-mass black holes, such as micro black holes, TeV-scale black holes, and potentially primordial black holes. The significance of this study lies in the unique intersection of astrophysics, particle physics, and quantum gravity that the analysis of accretion disks around quantum or small-mass black holes provides. This framework offers a means to test fundamental phenomena such as Hawking radiation and black hole evaporation, place constraints on primordial black holes as dark matter candidates, identify potential observational signatures of quantum gravity effects, and explore physics beyond the Standard Model. Additionally, we note that a similar quantum (string)-corrected black hole solution has been reported in Ref. \cite{Jusufi:2022uhk}. Both metrics exhibit a $\frac{1}{r^{4}}$ decay due to quantum corrections. In comparison, the Barbero--Immirzi parameter $\gamma$ in our model plays an analogous role to the parameter $\lambda$ introduced in the aforementioned reference.

According to Refs. \cite{akiyama2022first,doeleman2008event,raymond2024first}, here we want to examine whether the LQG corrections explored in this study can lead to potentially observable consequences. We assess the magnitude of the LQG-induced modifications in accretion disk flux and effective temperature. We then compare these theoretical predictions with the observational precision of current very long baseline interferometry (VLBI) instruments, particularly the Event Horizon Telescope (EHT). Therefore, we can define the relative shifts in flux and temperature as
\begin{equation}\label{delfrteffr}
\begin{split}
& \delta\mathcal{F}(r):=\frac{\mathcal{F}^{(\mathrm{LQG})}(r)-\mathcal{F}^{(\mathrm{Sch})}(r)}{\mathcal{F}^{(\mathrm{Sch})}(r)}\,,\quad\\
& \delta T_{eff}:=\frac{T_{eff}^{(\mathrm{LQG})}-T_{eff}^{(\mathrm{Sch})}}{T_{eff}^{(\mathrm{Sch})}}\,,
\end{split}
\end{equation}
where LQG refers to $\alpha\neq 0$ case and Sch refers to Schwarzschild case with $\alpha=0$. However, EHT does not directly observe flux or temperature at the accretion disk surface. Instead, it reconstructs intensity (brightness) profiles and image morphology from mm-wavelength VLBI data. The key characteristics of EHT data include an angular resolution of approximately $20-25\, \mu as$, an observed wavelength of $1.3\, mm\, (230\,\, GHz)$, and an image precision of about $10-20\%$ in brightness temperature at best.

The observed brightness temperature, which EHT reconstructs from data defined using observed specific intensity $I_{\nu}(\nu)$ is as follows
\begin{equation}\label{obbrte}
T_{b}(\nu)=\frac{c^{2}}{2k_{B}\nu^{2}}I_{\nu}(\nu)\,.
\end{equation}
The observed brightness temperature $T_{b}(\nu)$ is not the same as $T_{eff}$, but for a blackbody spectrum. However, if the disk emits as a blackbody (as assumed in the Novikov--Thorne model), then at long wavelengths or Rayleigh--Jeans limit (EHT's frequency), we have $T_{b}(\nu)\approx T_{eff}$. Therefore, we can quantify the relative shifts in flux and temperature to see if they in the image precision of EHT.

The values of $\delta\mathcal{F}(r)$ and $\delta T_{eff}$ for different values of $\gamma$ near the corresponding ISCO radii are collected in Table \ref{Table4}.
\begin{table}[hbt!]
  \centering
  \caption{\label{Table4}\small{\emph{Values of the relative shifts in flux and temperature near ISCOs of the loop quantum black hole arranged for the different values of $\gamma$.}}}
  \begin{tabular}{|p{1.5cm}|p{1.5cm}|p{1.5cm}|}
    \hline
    \multicolumn{3}{|c|}{Relative shifts} \\
    \hline
    \hline
    ${\gamma}$ & $\delta\mathcal{F}(r)$   & $\delta T_{eff}$ \\
    \hline
    \hline
    0.15 & 5\% & 5\% \\
    \hline
    0.21 & 14\% & 13\% \\
    \hline
    0.2375 & 20\% & 18\% \\
    \hline
    \end{tabular}
\end{table}
These relative shifts in flux and temperature of the loop quantum black hole become increasingly significant near the ISCO radius. Given that the EHT reports brightness calibration uncertainties on the order of $10\%–20\%$ for Sgr A* \cite{akiyama2022first}, we conclude that the predicted LQG-induced modifications to the disk flux and temperature may lie at the threshold of detectability with observational capabilities of EHT. This suggests potential relevance for future precision measurements of thermal disk spectra or reconstructed brightness distributions.

\section{Conclusions}\label{sec06}

In this study, we investigated motion of neutral and electrically charged particles, circular orbits and radiation properties of the accretion disk around the loop quantum black hole, as accretion disk radiation can be influenced by LQG parameter. The theoretical study of accretion disk radiation around a loop quantum black hole is considered to be a powerful test of its useful properties and astrophysical observations. We discovered that increasing the parameter $\gamma$ increases the effective potential of neutral particles in the loop quantum black hole, exactly the same for electrically charged particles far away from the source. Furthermore, we found that increasing the parameter $\gamma$ in the LQG black hole model  decreases the neutral test particle's specific energy, angular momentum and its angular velocity. To investigate the dynamics of electrically charged particles in the vicinity of a loop quantum black hole, we consider a background magnetic field that is static, axisymmetric, and asymptotically homogeneous. Within this framework, a Wald-like electromagnetic four-potential is employed, which includes the quantum corrections introduced by LQG. This potential approximately satisfies the covariant Maxwell's equations up to first order in the loop quantum parameter $\alpha$. Furthermore, we analyze the effective potential experienced by the charged test particle and compute the numerical values of the ISCO radii for the charged particles around the loop quantum black hole spacetime. Our findings indicate that increasing loop quantum parameter diminishes the ISCO radius of test particles, whether neutral or charged. Further, we investigated radiative efficiency and presented our findings in Table \ref{Table3}. Radiative efficiency rises as loop quantum parameter $\gamma$ grows.

Finally, we considered the accretion disk around the black hole as a primary source of information associated with the surrounding spacetime geometry and its nature in LQG. To provide valuable insights into the unique properties of the deformed Schwarzschild black hole inspired by LQG, we studied the influence of the loop quantum correction parameter on the accretion disk radiative properties, such as the flux of electromagnetic radiation, the temperature, differential luminosity, and the spectral luminosity of the disk. As a result the luminosity spectrum exhibits a blackbody-like distribution, characterized by a peak at higher frequencies. However, in the low-frequency regime, the spectral profiles corresponding to different values of $\gamma$ become nearly indistinguishable, making it difficult to differentiate between them. Interestingly, we found that the curves of these accretion disk radiation quantities shift upwards toward their bigger values as a result of the rise in the value of the loop quantum correction parameter $\gamma$, resulting in these quantities decreasing compared to the Schwarzschild and RN black hole cases in Einstein gravity. Based on the results, it was observed that the accretion disk around the loop quantum black hole can retain the radiative energy to prevent the electromagnetic radiation from emitting on the disk surface as a consequence of the influence of the loop quantum correction parameter $\gamma$.

\begin{acknowledgments}

The Work of K. Nozari and S. Saghafi is supported financially by the INSF of Iran under the grant number $4038520$. The authors want to express their gratitude to the respected referee for carefully reading the manuscript and the insightful comments.

\end{acknowledgments}

\appendix

\section{Delving into Satisfaction of Maxwell's Equations}\label{app1}

\subsection{The ordinary electromagnetic 4-potential}

We adopt a simple and exact configuration by assuming a purely azimuthal (Wald \cite{wald2010general}) electromagnetic 4-potential
\begin{equation}\label{ganefp11}
A_{\mu}=\left(0, 0, 0, \frac{B}{2}r^{2}\sin^{2}\theta\right)\,,
\end{equation}
where $B$ is the strength of the weak magnetic field. This corresponds to a uniform magnetic field aligned along the $z$-axis in the asymptotic limit, and apparently is an exact solution to Maxwell's equations. We now plan to verify whether the Maxwell's equations are satisfied by the simple electromagnetic 4-potential \eqref{ganefp11} in the spacetime of the deformed Schwarzschild black hole inspired by LQG. The Maxwell's equations in the spacetime without possible sources are as follows
\begin{equation}\label{mecs}
\nabla_{\mu}F^{\mu\nu}=\frac{1}{\sqrt{-\bar{g}}}\partial_{\mu}\left(\sqrt{-\bar{g}}F^{\mu\nu}\right)=0\,,
\end{equation}
where $\sqrt{-\bar{g}}=r^{2}\sin\theta$ in which $\bar{g}=g_{tt}g_{rr}g_{\theta\theta}g_{\varphi\varphi}$ is the metric determinant of the loop quantum black hole described in Eq. \eqref{Met}, and $F_{\mu\nu}=\partial_{\mu}A_{\nu}-\partial_{\nu}A_{\mu}$ is the electromagnetic field tensor.

Inserting Eqs. \eqref{Met} and \eqref{ganefp11} into Eq. \eqref{mecs} leads to the following result
\begin{equation}\label{mecs1}
\nabla_{\mu}F^{\mu\nu}=-\frac{3\alpha BM^{2}}{r^{6}}\neq 0\,,
\end{equation}
which clearly shows that the simple electromagnetic 4-potential \eqref{ganefp11} does not satisfy the Maxwell's equations in the deformed Schwarzschild black hole inspired by LQG with $\alpha\neq 0$. Therefore, we can conclude that generally the Maxwell's equations does not hold independently of the form of $f(r)$ in any spherically symmetric spacetime. Consequently, we may accept that no non-trivial exact electromagnetic field exists without solving Maxwell's equations with full LQG corrections. We have to propose an alternative electromagnetic 4-potential including $\alpha$ term, which approximately solves the Maxwell's equations in the spacetime.

\subsection{The electromagnetic 4-potential inspired by LQG}

Inspired by Wald's solution \cite{wald2010general}, we propose the following general ansatz for the electromagnetic 4-potential
\begin{equation}\label{ganefp}
A_{\mu}=\left(0, 0, 0, \frac{B}{2}\Psi(r)\sin^{2}\theta\right)\,,
\end{equation}
where $\Psi(r)$ is an unknown function of $r$, which we want to determine it such that Maxwell's equations are satisfied. Inserting Eqs. \eqref{Met} and \eqref{ganefp} into Eq. \eqref{mecs} leads to the following result
\begin{equation}\label{maxinin}
\nabla_{\mu}F^{\mu\nu}=\frac{B}{r^{2}}\left(\frac{1}{2}\left(f(r)\Psi'(r)\right)'-\frac{1}{r^{2}}\Psi(r)\right)=0\,.
\end{equation}
As a result of rearranging Eq. \eqref{maxinin}, we obtain the primary condition required to satisfy Maxwell's equations by the considered ansatz for the electromagnetic 4-potential
\begin{equation}\label{mancon}
\left(f(r)\Psi'(r)\right)'=\frac{2}{r^{2}}\Psi(r)\,.
\end{equation}

Let's consider the following assumption
\begin{equation}\label{Psiasum}
\Psi(r)=r^{2}+\frac{\alpha}{r^{2}}\,.
\end{equation}
This assumption is quite elegant; it reduces to the standard Wald solution \eqref{ganefp11} when $\alpha\to 0$ and introduces a correction that is proportional to $1/r^{2}$, as expected from quantum corrections. By taking into account the assumption \eqref{Psiasum}, the left-hand side of Eq. \eqref{mancon} becomes 
\begin{equation}\label{leftside}
1+\frac{3\alpha}{r^{4}}-\frac{3\alpha M^{2}}{r^{4}}-\frac{8\alpha M}{r^{5}}+\frac{7\alpha^{2}M^{2}}{r^{8}}\,,
\end{equation}
while the right-hand side of Eq. \eqref{mancon} becomes 
\begin{equation}\label{rightside}
1+\frac{\alpha}{r^{4}}\,.
\end{equation}
Thus, this ansatz meets the covariant Maxwell's equations up to $\mathcal{O}(\alpha)$, indicating consistency with minor quantum corrections from LQG, though it does not fully satisfy Maxwell's equation. Let's quantify the size of the residual Maxwell's equations as follows
\begin{equation}\label{resmaxequ}
\delta(r):=\left(f(r)\Psi'(r)\right)'-\frac{2}{r^{2}}\Psi(r)\,.
\end{equation}
Using \eqref{resmaxequ}, one can deduce that
\begin{equation}\label{evalresu}
\mathrm{Relative\,Error}:=\left|\frac{\delta(r)}{\frac{2}{r^{2}}\Psi(r)}\right|\ll 1\%\,,
\end{equation}
is always satisfied for the range of $r>4GM$ with the given values of $\alpha$ in our setup. Therefore, the approximate ansatz \eqref{Psiasum} is extremely close to solving the exact Maxwell's equations. Consequently, the impact on observables like energy, angular momentum, or ISCO is also negligible. This makes it clear that the approximation is under control and physically insignificant for the observables of interest.

Consequently, we take into account the Wald-like electromagnetic 4-potential
\begin{equation}\label{wtemfpa1}
A_{\mu}=\left(0, 0, 0, \frac{B}{2}\left(r^{2}+\frac{\alpha}{r^{2}}\right)\sin^{2}\theta\right)\,,
\end{equation}
which approximately satisfies the covariant Maxwell's equations up to $\mathcal{O}(\alpha)$.

\end{document}